\documentclass[11pt, a4paper]{article}

\pdfoutput=1

\usepackage{amsmath}
\usepackage{amsfonts}
\usepackage{amssymb}
\usepackage{cite}
\usepackage{multicol,multirow}
\usepackage{graphicx, physics, subcaption}
\usepackage{graphics, epsfig}
\usepackage{epsf}
\usepackage{epstopdf}
\usepackage[usenames,dvipsnames]{color}

\usepackage{colortbl}
\definecolor{summersky}{cmyk}{0.71,0.33,0,0.14}
\definecolor{flamingo}{cmyk}{0,0.51,0.71,0.14}
\definecolor{rp}{cmyk}{0.2, 1, 0.6, 0}
\definecolor{pacificblue}{cmyk}{0.95,0.3,0, 0.19}
\definecolor{gray60}{cmyk}{0.4,0.4,0,0.8}

\numberwithin{equation}{section}

\usepackage[colorlinks,bookmarks=false, hypertexnames=true]{hyperref}
\hypersetup{
	breaklinks=true,
	pdfstartview={FitH},    
	colorlinks=true,       
	linkcolor=blue,          
	citecolor=red,        
	filecolor=magenta,      
	urlcolor=red,           
	anchorcolor=green,      
	linktocpage=true
}

\newcommand{\nc}{\newcommand}

\nc{\ba}{\begin{eqnarray}}
\nc{\ea}{\end{eqnarray}}

\nc{\calR}{{\cal{R}}}
\nc{\calP}{{\cal{P}}}
\nc{\cN}{ {\cal{N}} }

\def\calM{{\cal M}}

\begin{document}

\vspace{5mm}
\vspace{0.5cm}
\begin{center}

\def\thefootnote{\fnsymbol{footnote}}

{\bf   Regular Black Holes from Anisotropic Source 
with Hydrodynamic Equation of State  }
\\[0.5cm]

{  Hassan Firouzjahi\footnote{firouz@ipm.ir }}

{\small \textit{School of Astronomy, Institute for Research in Fundamental Sciences (IPM) \\ P.~O.~Box 19395-5746, Tehran, Iran }}\\

\end{center}

\vspace{.8cm}

\hrule \vspace{0.3cm}


\begin{abstract}

We study regular black hole solutions sourced by an anisotropic energy momentum tensor. It is well known that the geometry of the interior of a spherically symmetric regular black hole approaches the dS metric.  
Having decomposed the energy momentum tensor into its isotropic and anisotropic components, we assume a hydrodynamic equation of state, $P= P(\rho)$, for the pressure, and look for spherically symmetric, regular black hole solutions.  We consider different forms of $ P(\rho)$ which yield the previously known regular black hole solutions, as well as various new metrics.  We show that the profile of $ P(\rho)$ has a root and a maximum as it approaches $0^+$ at large distances.  Consequently, the square of 
the sound speed of perturbations, $c_s^2$, changes sign at the point where $P$ reaches its maximum, indicating a potential hydrodynamic instability. In addition, imposing the subluminal bound on $c_s$ puts strong constraints on  the model parameters, excluding models in which the energy density has an exponential fall off.  We establish a universal hierarchy among the relative positions at which the strong energy condition is violated, at which $P$ has its root, and at which $P$ attains its maximum.

\end{abstract}
\vspace{0.5cm} \hrule
\def\thefootnote{\arabic{footnote}}
\setcounter{footnote}{0}
\newpage
\section{Introduction}

Singularities of black holes (BHs) are  challenges for theoretical general relativity (GR). The existence of singularities in GR are unavoidable under some certain physical assumptions on the  energy momentum tensor and the global properties of spacetime \cite{Penrose:1964wq, Hawking:1970zqf, Hawking:1973uf}. Typically, one expects that curvature invariants and the energy density and pressure to blow up at the center of BHs. This indicates that the classical GR is inadequate deep in the interior of BHs and some new physics such the quantum effects may be required to bypass the singularities in the interior o BHs.

There have been active interests in literature to construct regular BHs, i.e. BH solutions which do not suffer from interior singularities, see  \cite{Ansoldi:2008jw, Frolov:2016pav, Maeda:2021jdc} for some reviews. 
To achieve this, some parts of the theorems leading to singularities have to be violated. One approach is to violate the strong energy condition (SEC) which seems less dramatic and can be realized in various field theory setups while the weak energy condition (WEC) still does hold.  The first concrete example of regular BHs was proposed by Bardeen 
\cite{Bardeen68} without specifying the matter content yielding to this metric. Later many regular BHs were constructed  
\cite{ Dymnikova:1992ux, Dymnikova:2004zc, Dymnikova:2003vt, Dymnikova:2001fb, Mbonye:2005im, Hayward:2005gi, Fan:2016hvf}.  All these regular BHs have the common property that the interior of BH is replaced by a dS spacetime where it is well known that the SEC is violated. The spacetime approaches the Schwarzschild solution on large distance with a finite total mass. Typically, the model is identified by two parameters, $\ell, m$ in which $\ell$ is a length scale where the new physics enter and the violation of SEC becomes relevant. The mass scale $m$ measures the total mass as measured by an observer in far distances. It is realized that these regular BHs can be constructed within the non-linear electrodynamics with a non-linear form of the Lagrangian \cite{Ayon-Beato:1998hmi, Ayon-Beato:1999qin, Ayon-Beato:1999kuh, 
Bronnikov:2000vy, Bronnikov:2017sgg, Bronnikov:2022ofk, Dymnikova:2004zc, Balart:2014cga, Maeda:2010qz, Fan:2016hvf}. We comment that the idea of replacing the singularity of the interior region by a dS spacetime such as via matching condition or the proposal of maximum curvature hypothesis 
was studied in  other works as well \cite{Sakharov:1966aja, Poisson:1988wc, Frolov:1989pf, Frolov:1988vj, Firouzjahi:2016nle}. 

In this work we study regular spherically symmetric  BHs from anisotropic fluid  with a known equation of state (e.o.s.). In previous literature the regular BH solutions are constructed via inverse engineering. More specifically, one assumes a specific form of of the energy density profile $\rho(r)$ which satisfies basic criteria both at the center $r=0$ and on far distances $r\rightarrow \infty$. Then one plugs this proposed form of $\rho(r)$ into the Einstein field equations and obtain the metric function and the remaining components of the energy momentum tensor. Here we consider a different approach in which we specify the e.o.s. of the pressure $P=P(\rho)$ and then look for 
the consistent solutions of regular BHs. After discussing some general conclusions learned from this analysis, we consider some specific examples of $P(\rho)$ which yield not only the  previously known regular BH solutions but also some new metrics.

\section{Black Hole from Anisotropic Fluid}
\label{setup}

Here we present the field equations for BH sourced by  an anisotropic fluid.

We consider the spherically symmetric static BH 
with the following line element in radial coordinate,  
\ba
\label{metric-BH}
ds^{2 } = -A(r) dt^2+ \frac{ dr^2}{B(r)}   + r^{2 } d \Omega^{2} \, ,
\ea
in which $d \Omega^{2}= d \theta^2 + \sin^2\theta d\phi^2$ represents the 
line element of a unite two-sphere. Here we started   with the general case where there are two metric functions $A(r)$ and $B(r)$. However,  we will eventually  study the case with $A(r)= B(r)$ where it is also shown under what condition on the source of energy momentum tensor this is a consistent solution.  

The source of energy momentum tensor is an anisotropic fluid in which $T_{\mu \nu}$ takes the following form \cite{Ellis:1998ct}, 
\ba
\label{T-general}
T_{\mu\nu} = (\rho + P) u_{\mu} u_{\nu} + P g_{\mu \nu} + q_{\mu} u_{\nu} +  q_{\nu} u_{\mu} + \pi_{\mu \nu} \, ,
\ea
subject to  the conditions,
\ba
u^{\mu}u_{\mu} =-1, \quad 
q_{\mu} u^{\mu}=0, \quad \pi^{\mu}_{\mu} =0 ,\quad \pi_{\mu \nu} = \pi_{\nu \mu} , \quad \pi_{\mu \nu} u^{\mu}=0 \, ,
\ea
in which $\rho$ is the energy density, $P$ is  the isotropic pressure, $u^{\mu}$ is the comoving four velocity associated to the fluid, $\pi_{\mu \nu}$ is the anisotropic pressure and $q^{\mu}$ is the heat conduction which represents the energy flux relative to $u^{\mu}$. We consider the simplified setup in which  there is no heat flow for the fluid so $q_{\mu}=0$. In our analysis below  we consider a single fluid but the extensions to multiple fluids  can be considered as well.

From the symmetry of the background, the comoving velocity $u^\mu$ is given by,  
\ba
u^{\mu} = (A^{-1/2}, 0, 0, 0) \, .
\ea
With the assumption of no heat flow, $T^{\mu}_{\nu}$ takes the diagonal form 
\ba
T^{\mu}_{\nu} = \mathrm{diag} \left(   -\rho, P + \pi^{1}_{1}, P+ \pi^{2}_{2} , P+ \pi^{3}_{3}
\right) \, .
\ea
Imposing the symmetry condition along the spherical directions 
$\pi^{2}_{2}= \pi^{3}_{3}$ and the traceless condition, $\pi^{\mu}_{\mu}=0$, we obtain \cite{Firouzjahi:2024xbo}, 
\ba
\label{T-mu-nu}
T^{\mu}_{\nu} = \mathrm{diag} \left(   -\rho, P - 2 \Pi, P+\Pi , P+ \Pi \right) \, ,
\ea
where $\Pi\equiv \pi^{2}_{2}$ controls the anisotropic pressure. 

From the Einstein field equation $G_{\mu\nu}= 8 \pi G T_{\mu \nu}$, and working in convention where $G=1$, we obtain 
\ba
\label{A-eq}
r   B' + B - 1 =-8 \pi r^{2 }  \rho(r)  \, , \\
\label{A-B-eq}
\frac{B}{r} \left( \frac{ A'}{A} - \frac{ B'}{B}   \right) = 8 \pi\big( \rho + P - 2 \Pi \big) \, ,
\ea
in which a prime indicates the derivative with respect to $r$.

The remaining independent equation can be obtained from the energy conservation $\nabla_{\mu} T^{\mu \nu}=0$, yielding
\ba
\label{conservation-eq}
\big( P- 2 \Pi\big)' + \frac{A'}{2 A} \big( P- 2 \Pi + \rho) -\frac{6}{r} \Pi =0 \, .
\ea

As mentioned before, we are interested in solutions with one metric function as in Schwarzschild metric in which $A(r)= B(r)$. Looking at Eq. (\ref{A-B-eq}), this requirement immediately yields,
\ba
\label{Pi-eq}
\Pi = \frac{\rho+ P}{2} \, .
\ea
Intuitively speaking, the anisotropic pressure is the average of the energy density and the isotropic pressure. 

Plugging the above constraint into the energy conservation equation (\ref{Pi-eq}) we obtain the following simple equation for the evolution of $\rho$,
\ba
\label{energy-eq}
 \rho' + \frac{3}{r} (\rho + P) =0 \, .
\ea
This is a simple and yet powerful equation. Interestingly, it looks like the energy conservation in FLRW cosmology if one replaces $r$ by $t$ and the radial profile $r^2$ by the scale factor $a(t)^2= t^2$. This analogy becomes more clear if one considers the interior of the BH where the roles of the $(r, t)$ coordinates are switched. More specifically, in the interior of BH, $r$ becomes a timelike coordinate and the spacetime represents an anisotropic cosmological background known as the Kantowski-Sachs \cite{Kantowski:1966te} spacetime. 
Correspondingly, the energy conservation equation (\ref{energy-eq}) takes the cosmological form $\dot \rho+ 3 H (\rho+ P)=0$ in which $H= \dot a(t)/a(t)=1/t$ is the isotropic Hubble rate (contraction) in the interior of BH. 

Finally, with $B(r)= A(r)$, the metric function from Eq. (\ref{A-eq}) can be obtained via an integration over $\rho(r)$. Defining the mass function via 
\ba
\label{mass-eq}
A(r)= 1- \frac{2 M(r)}{r},
\ea
we obtain 
\ba
\label{M(r)}
M(r)=   4 \pi \int_0^r dx x^2 \rho(x) \, ,
\ea
in which we have assumed $M(0)=0$ so there is no singularity at the center 
of BH. In general, if one considers singular BHs as well, then $M(0)\neq 0$ as in Schwarzschild solution. Having said this,  the condition $M(0)=0$ imposes a strong fine-tuning as there is no apparent dynamical mechanism to prevent $M(0)\neq 0$. 

Before we proceed, we comment that usually in the literature 
the energy momentum tensor  associated to the anisotropic fluid is written as,
\ba
\label{T-lit}
T^{\mu}_{\nu} = \mathrm{diag} \left(   -\rho, P_r, P_t, P_t \right) \, ,
\ea
in which $P_r$ and $P_t$ are the radial and the tangential pressures respectively.  Then, the condition (\ref{Pi-eq}) is translated into $P_r= -\rho$ and the energy conservation equation (\ref{energy-eq}) is used to solve the tangential pressure as $P_t(r)= -\rho-r \rho'/2 $. While this is algebraically consistent, but it may hide the correct interpretations when reading the hydrodynamic equation of state. We demonstrate this in a simple example involving the Maxwell field in Appendix  \ref{Maxwell-example}. 

\section{Conditions for Regular Black Hole Solutions}
\label{Reg-BH-cond}

Here we present our assumptions and the general requirements for the existence of the regular BHs. Most of the results presented here are known
in previous literature, see for example \cite{Maeda:2021jdc, Ansoldi:2008jw} for a review. But here we present the results with a different perspective, with emphasis on our choice of the form of the anisotropic $T^{\mu \nu}$ given in Eq. (\ref{T-general}) and the roles of the equation of state for the isotropic pressure $P=P(\rho)$. Along the way we present some new results as well.  

We impose the following physically well-motivated assumptions on the  energy momentum tensor sourcing the regular BH solutions:
\vspace{0.3cm}

{\bf (a)}- The weak energy condition (WEC) does hold throughout with $\rho\ge0$ and  $\rho+P \ge 0$. Combined with Eq. (\ref{energy-eq}), the WEC assumption  implies that  $\rho(r)$ is a decreasing function,
 $\rho'(r)\leq 0$. 
 \vspace{0.3cm}
 
{\bf (b)}- Both $\rho$ and $P$ are finite in all spacetime, and in particular at the center $r=0$. These may be viewed as the physical requirements for the existence of the regular BHs as consistent solutions of the Einstein field equations. 
\vspace{0.3cm}

{\bf (c)}- The finiteness of the total mass,
\ba
\label{M-total}
{\cal M} \equiv M(\infty)=   4 \pi \int_0^\infty dx\,  x^2 \rho(x)  < \infty  \, ,
\ea
so the spacetime is asymptotically flat, resembling  the Schwarzschild  metric on large distances. The finiteness of $\calM$ requires that $\rho(r)$ falls off faster than $r^{-3}$ on large distances. 
\vspace{0.3cm}

Having the above three assumptions imposed, let us start with Eq. (\ref{energy-eq}) to solve $P(r)$ in terms of $\rho(r)$, obtaining, 
\ba
\label{P-eq}
P(r)= -\frac{1}{3 r^2} \big( \rho r^3\big)' \, .
\ea
Requiring  $\rho(r)$ to be finite at $r=0$, we conclude that $P<0$ near the center. On the other hand, demanding that $\calM$ to be finite requires that $\rho r^3$ to be a decreasing function so $P>0$ on large distances. The important conclusion is that $P$ switches the sign from a negative value near the canter towards a positive value in asymptotically large distance. As a result, $P(r)$ has at least one zero and one maximum in the range $0< r < \infty$. Now consider the simple cases  where $P(r)$ has only one root and one maximum (as in well-known examples of regular BHs solutions).  Denoting the position of the maximum of $P$ by $r_m$ and the position where $P$ vanishes by $r_*$, from the 
 asymptotic structure of $P(r)$ outlined above we conclude that 
$r_* < r_m$. 

Consider the Taylor expansion $\rho(r)= \rho_0 + \rho_1 r+ \rho_2 r^2+...$ and $P= P_0 + P_1 r + P_2 r^2+...$ near the origin. This is allowed since we have assumed that $\rho$ and  $P$ are finite at the center. Plugging the above expansion into Eq. (\ref{P-eq}), we conclude that $P_0= -\rho_0$, $P_1= -4\rho_1/3$ and so on. In particular, from the fact that $P_0 = -\rho_0$ we conclude that the spacetime near the center is a dS type. This is a known result in which the regularity of the solution requires that the spacetime near the canter approaches a dS spacetime \cite{Dymnikova:1992ux}. However, to prove the regularity of spacetime mathematically,  we also have to look at the curvature invariants such as the Ricci scalar, Kretschmann  invariant etc to examine the finiteness of the geometric quantities as well. A careful investigation demonstrates that indeed the spacetime should approach a dS limit near the center of regular BH \cite{Maeda:2021jdc}.

On the other hand, consider the large distance limit where  $\rho(r)$ is required to  fall off faster that $r^{-3}$. Let us first consider a power law fall off with $\rho \propto r^{-(3 + \alpha)r}$ with $\alpha >0$. Plugging this value of $\rho$ in Eq. (\ref{P-eq}) we obtain $P = \frac{\alpha}{3} \rho$ for $r\rightarrow \infty$.  This yields the linear e.o.s. $P=w \rho$ with $w=\alpha/3$.  As the second example, now suppose we have an exponential fall off 
 $\rho(r) =\rho_\infty e^{- \beta r}$ with $\beta >0$. Then, from Eq. (\ref{P-eq}) we obtain $P= -\frac{\rho}{3} \ln\big(\frac{\rho}{\rho_\infty} \big)  $. This suggests a non-linear e.o.s. on large distances. 
 
 A non-trivial conclusion from the existence of regular BH solution is that the strong energy condition (SEC) should be violate somewhere \cite{Dymnikova:1992ux, Elizalde:2002yz, Zaslavskii:2010qz, Maeda:2021jdc}. The SEC requires that  $\rho+ 3 P \ge 0$ so to have a regular BH solution $\rho+ 3 P$ should changes sign. To see this, let us use the value of $P$ obtained in Eq. (\ref{P-eq}) to write,
 \ba
 \label{SEC}
 \rho+ 3 P= -\frac{1}{r} \big( \rho r^2)' \, .
 \ea
 Near the center $\rho\rightarrow \rho_0$ and $P\rightarrow - \rho_0$ so we obtain 
 $\rho+ 3 P \rightarrow \rho_0+ 3 P_0 = -2 \rho_0<0$ so indeed deep 
 in the interior region SEC is violated. On the other hand, for large distance we know that $\rho r^3$ is a decreasing function so $\rho r^2$ is a decreasing function as well. Correspondingly $(\rho r^2)'<0$ and 
 $\rho+ 3 P >0$ on large distances. As a result, we conclude that $\rho+ 3 P$ should switch sign in the intermediate region.

One can naturally compare the position where the SEC is violated, denoted by $r_v$, with $r_*$ where $P$ vanishes. At the point $r= r_*$, we obtain $\rho(r_*) + 3 P(r_*) = \rho(r_*) >0$ so the SEC is not violated at $r_*$. Since the sign of 
 $\rho+ 3 P$ is positive in far distances and negative close to the center, we conclude that $r_v < r_*$. In other words the root of $P$ is further away from the point where the SEC is violated. As a result, we have the hierarchy $r_v< r_* < r_m$.  Indeed, we will demonstrate in the examples presented in next sections that $r_v$ is mostly close to the center of the BH and the above hierarchy always hold.   
 
\begin{figure}[t]
\begin{center}
	\includegraphics[scale=0.45]{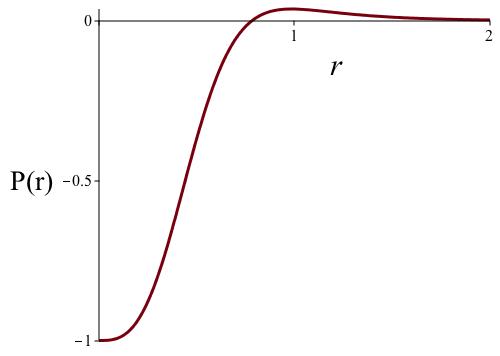}
	\hspace{1cm}
	\includegraphics[scale=0.45]{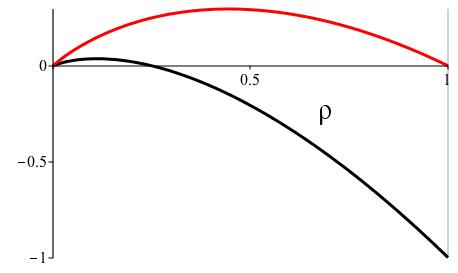} 
	\end{center}
\caption{ Left:  $P(r)$ as a function of $r$. We observe the general behaviour that $P<0$ at the center of BH while approaching $0^+$ on large distances. $P(r)$ has a maximum at $r_m$ and a root at $r_*$.  Right:  $P(\rho)$ (lower black curve) and $f(\rho)\equiv\rho+P(\rho)$ (upper red curve) in terms of $\rho$.  It is observed  that $f(\rho)\ge 0$ due to WEC and it has roots at $\rho=0, 1$.  Both plots are for the Hayward  BH with  $\rho_0=1$.
\label{P-plot}
}
\end{figure}

For a general view of $P(r)$ and $P(\rho)$ see the plots in Fig. \ref{P-plot}.
 These plots are obtained for the Hayward regular BH \cite{Hayward:2005gi} (reviewed in next section) in the unit $\rho_0 \equiv \rho(r=0)=1$. In the left panel we observe that $P(0)=-\rho_0<0$ while it approaches $0^+$ on large distances. It has a toor at $r_*$ and a maximum at $r_m$. In the right panel $P(\rho)$ and $f(\rho)\equiv \rho+P(\rho)$ are plotted. In particular, we observe that $f(\rho)\ge 0$ due to WEC. The existence of two roots for $f(\rho)$ at the two endpoints is the basic requirement for the existence of the regular BH solution. 

An important conclusion from the above discussions is that the square of the sound speed of perturbations, 
\ba
\label{cs-def}
c_s^2 \equiv  \frac{\partial P}{\partial \rho} \, ,
\ea
changes sign at the point where $P(\rho)$ reaches its maximum (a similar conclusion was also reached in \cite{Perez:2014oea} for the regular BH setup proposed in  \cite{Mbonye:2005im}).  Specifically, from the general behaviour of $P(\rho)$ reviewed above we conclude that 
$c_s^2 >0$  for $r>r_m$ corresponding to $\rho < \rho(r_m)$. On the other hand, for  $r<r_m$ and $\rho > \rho(r_m)$ (i.e. near the interior of BH)  we have $c_s^2 <0$. This indicates a possible hydrodynamic instability. Indeed the quasi normal modes (QNMs) perturbations of a number of regular BHs were studied where they are shown to be stable \cite{Moreno:2002gg, Breton:2005ye, 
Flachi:2012nv, Chaverra:2016ttw, Wu:2018xza, Nomura:2020tpc}. This may be contradictory to our conclusion that $c_s^2$ become negative in some regions. The paradox may be resolved noting that the analysis in these papers were concerned about the QNMs which are defined for the exterior region of BHs. However, we see that $c_s^2$ switches sign mostly in the interior region (see the end of subsection \ref{reg-hyd} for further discussions) where also it is well understood that the inner horizon suffers from mass inflation instability \cite{Poisson:1989zz, Poisson:1990eh}. Having said this, it is an interesting question to extend the perturbation analysis of \cite{Moreno:2002gg, Breton:2005ye, 
Flachi:2012nv, Chaverra:2016ttw, Wu:2018xza, Nomura:2020tpc} to the interior region and check whether or not there is a link between $c_s^2$ switching sign and the (in)stability of the interior spacetime.

To have regular BH solution we require that the metric function $A(r)$ to have at lease two roots \cite{ Frolov:2016pav, Maeda:2021jdc}. The larger root at $r= r_+$ represents the outer horizons (the  event horizon) while the smaller toot at $r= r_-$ is the position of the inner horizon (the Cauchy horizon). The conditions of the existence of the horizons  impose constraints on the values of the model parameters.

In usual treatment of regular BHs in literature, one employs an inverse engineering method that first one choses a desirable profile of $\rho(r)$. Then plugging $\rho(r)$ in  Eq. (\ref{M(r)}) the mass function for the component of the metric function is obtained. Finally, the tangential pressure $P_t$ is obtained from the energy conservation equation (written in terms of $P_t$ instead of  $P$). However, a more physically reasonable approach would be to specify the matter content of the theory (i.e. the field theory behind the model) and then solve the Einstein field equations to obtain 
$\rho(r), P(r)$ and $M(r)$. Usually this is a non-trivial task. Having said this, it turns out that all know solutions of the regular BHs can be realized
within  non-linear electrodynamics (NEDs) with the Lagrangian density ${ L}({\cal F})$ in which ${\cal F} \equiv F_{\alpha \beta} F^{\alpha \beta}$ and $F_{\alpha \beta}$ is the field strength of the $U(1)$ gauge field.   
One drawback of employing 
NEDs to obtain the regular BHs is that the Hamiltonian in these solutions 
does not reduce to the free Maxwell theory in the IR limit. 

Motivated by the above discussions, in this work we employ an intermediate approach compared to the above two approaches. We assume specific  non-linear hydrodynamic equation of state $P= P(\rho)$ and then solve the energy conservation equation (\ref{energy-eq}) to find $\rho= \rho(r)$. Having obtained $\rho(r)$ one can use Eq. (\ref{M(r)}) to find the mass function of the regular BH as usual. One shortcoming of this hydrodynamic approach is that in general we do not know the field theory behind the model yielding the non-linear e.o.s. $P=P(\rho)$.

\section{Regular Black Holes}
\label{reg-BH}

Here we consider some well-know cases of regular BHs which are proposed in the literature and obtain  the corresponding equation of state $P=P(\rho)$. In addition, we study other physical questions such as the position  where the SEC is violated $(r_v)$, the position where $P$ changes sign $(r_*)$, the position of its maximum value $(r_m)$ and the square of the sound speed $c_s^2$.

Among the well-studied models of regular BHs are the Bardeen spacetime \cite{Bardeen68}, Hayward spacetime \cite{Hayward:2005gi}, Dymnikova I and II spacetimes \cite{Dymnikova:1992ux, Dymnikova:2004zc}, Fan-Wang spacetime \cite{Fan:2016hvf}, for a review see \cite{Maeda:2021jdc}. 

With the mass function defined in Eq. (\ref{mass-eq}), and following the notation of \cite{Maeda:2021jdc},  the forms of $M(r)$ for these metrics are given as follows,
\ba
\label{Bardeen}
M(r) &=&  \frac{m r^3}{(r^2 + \ell^2)^{3/2}} \hspace{4cm} (\mathrm{Bardeen})\\
\label{Hayward}
M(r) &=&  \frac{m r^3}{ r^3 + 2 m \ell^2} \hspace{4cm} (\mathrm{Hayward})\\
\label{Fan-Wang}
M(r) &=&  \frac{m r^3}{ ( r + \ell)^3} \hspace{4.3cm} (\mathrm{Fan-Wang})\\
\label{DymnikovaI}
M(r) &=& \frac{2 m}{\pi} \Big[ \arctan\big( \frac{r}{\ell}\big) - \frac{\ell r}{r^2+ \ell^2}
\Big]  \quad \quad (\mathrm{Dymnikova~ I}) \\
\label{DymnikovaII}
M(r) &=& m ( 1- e^{- r^3/\ell^3})  \hspace{3cm} (\mathrm{Dymnikova~ II}) \, .
\ea
These spacetimes have two parameters $(m, \ell)$ in which the length scale parameter $\ell$ measures the scales inside the BH where the modification from the Schwarzschild metric becomes significant. In addition, the mass scale $m$ is defined such that $\mathrm{lim}_{r \rightarrow \infty} M(r)= m$.
In our notation, $m= \calM$.  Near the center $r\rightarrow0$, these spacetimes approach a dS metric in which $M(r) \propto r^3$. For example, for the Bardeen and Hayward metrics, one obtains,
\ba
m(r)\simeq \frac{m}{\ell^3} r^3 -\frac{3 m}{2 \ell^5} r^5 \, , 
\quad \quad (\mathrm{Bardeen},~ r \rightarrow 0) \, ,
\ea 
and
\ba
m(r)\simeq \frac{1}{2\ell^2} r^3 -\frac{1}{4 m \ell^4} r^6 \, 
\quad \quad (\mathrm{Hayward},~ r \rightarrow 0) \, .
\ea 
 All these metrics have two horizons, the outer horizon is an event horizon while the internal one is a Cauchy horizon. 


\subsection{hydrodynamic Equation of State}
\label{reg-hyd}

With the mass functions given in Eqs. (\ref{Bardeen})-(\ref{DymnikovaII}), we can construct $\rho(r)$ from $M'(r)$  in  Eq. (\ref{M(r)}) and then obtain $P(r)$ from  the energy Eq. (\ref{P-eq}). The last step is to eliminate $r$ in terms of $\rho$
to obtain  $P= P(\rho)$. We present the results for each case in turn.

\subsubsection{Bardeen BH}
For the Bardeen metric, we obtain 
\ba
\rho(r)= \frac{3 m \ell^2}{4 \pi (\ell^2+ r^2)^{5/2}} \, , \quad \quad
P(r)= \frac{\ell^2 m ( 2 r^2 - 3 \ell^2 )}{4 \pi (\ell^2 + r^2)^{7/2}} \, .
\ea
Eliminating $r$ from $\rho(r)$ and plugging the result into $P(r)$ we 
obtain $P(\rho)$ as follows,
\ba
\label{EOS-B}
P(\rho)= \frac{2 \rho}{3} \Big(  1- \frac{5}{2} \big(\frac{\rho}{\rho_0})^{\frac{2}{5}}\Big) \, ,
\ea
in which $\rho_0\equiv 3m/4 \pi \ell^3 $ is the  value of $\rho$ at the center. 
One can check that near  the center $P\rightarrow -\rho$ while for far distances it reaches a linear e.o.s with $P\rightarrow 2\rho/3$. 

Calculating the root of $\rho+ 3 P$, the root of $P$ and its maximum we obtain,
\ba
r_v= \frac{\sqrt6 \ell}{3} \, , \quad \quad
r_*=  \frac{\sqrt6 \ell}{2}  \, , \quad \quad
r_m= \frac{\sqrt{10} \ell}{2}  \, .
\ea
We see that  the hierarchy 
$r_v < r_* < r_m$ does hold as anticipated in  section \ref{Reg-BH-cond}. 

The sound speed $c_s^2$ is obtained to be,
\ba
c_s^2 &=&\frac{\partial P}{\partial \rho}= \frac{2}{3}- \frac{7}{3} \big(\frac{\rho}{\rho_0} \big)^{2/5} \, \nonumber\\
&=& \frac{2 r^2 - 5 \ell^2}{3 (r^2 + \ell^2) }  \, .
\ea
As expected $c_s^2$ changes sign at $r= r_m$,  with the asymptotic value 
$c_s^2 \rightarrow 2/3$ on IR limits (long distances).

\subsubsection{Hayward BH}
For the Hayward metric, we obtain 
\ba
\rho(r)= \frac{3 m^2 \ell^2}{2 \pi (2 m \ell^2+ r^3)^{2} }\, ,\quad \quad
P(r)=  \frac{3 m^2 \ell^2 ( r^3- 2 m \ell^2 )}{2 \pi (2 m \ell^2+ r^3)^{3} } \, .
\ea
Eliminating $r$ from $\rho(r)$ and plugging the result into $P(r)$ we 
obtain the corresponding e.o.s.,
\ba
\label{EOS-H}
P(\rho)= \rho \Big(  1- 2 \big(\frac{\rho}{\rho_0})^{\frac{1}{2}}\Big) \, ,
\ea
in which $\rho_0\equiv 3/8 \pi \ell^2 $. 
Near  the center $P\rightarrow -\rho$ while on large distances it reaches a linear e.o.s with $P\rightarrow \rho$. 

Calculating $r_v, r_*$ and $r_m$ we obtain,
\ba
r_v= (m \ell^2)^{\frac{1}{3}}\, , \quad \quad
r_*=  (2m \ell^2)^{\frac{1}{3}} \, , \quad \quad
r_m= (4m \ell^2)^{\frac{1}{3}}   \, .
\ea
As expected,  $r_v < r_* < r_m$.

The sound speed $c_s^2$ is, 
\ba
c_s^2 = 1- 3  \big(\frac{\rho}{\rho_0} \big)^{\frac{1}{2}} \, 
= \frac{r^3 - 4 m \ell^2}{ r^3 + 2 m \ell^2 }  \, ,
\ea
with a root at $r_m$ and the asymptotic value 
$c_s^2 \rightarrow 1$ on long distances.

\subsubsection{Fan-Wang BH}
For the Fan-Wang metric, we obtain 
\ba
\rho(r)= \frac{3 m \ell}{4 \pi ( \ell+ r)^{4} }\, , \quad \quad
P(r)=  \frac{ m \ell (r- 3 \ell) }{4 \pi ( \ell+ r)^{5} }\, .
\ea
From the above expressions, the e.o.s. is obtained to be, 
\ba
\label{EOS-H}
P(\rho)= \frac{\rho}{3} \Big(  1- 4 \big(\frac{\rho}{\rho_0})^{\frac{1}{4}}\Big) \, ,
\ea
in which $\rho_0\equiv 3m/4 \pi \ell^3 $. 
Near  the center $P\rightarrow -\rho$ while on large distances it reaches 
the relativistic e.o.s. $P\rightarrow \rho/3$. 

Calculating $r_v, r_*$ and $r_e$ we obtain,
\ba
r_v= \ell\, , \quad \quad
r_*=  3 \ell\, , \quad \quad
r_m= 4 \ell  \, .
\ea
As expected,  $r_v < r_* < r_m$.

The sound speed $c_s^2$ is, 
\ba
c_s^2 = \frac{1}{3}- \frac{5}{3}  \big(\frac{\rho}{\rho_0} \big)^{\frac{1}{4}} \, 
= \frac{r - 4 \ell}{ 3( r + \ell) }  \, ,
\ea
with a root at $r_m$ and the asymptotic value 
$c_s^2 \rightarrow 1/3$ on long distances.

\subsubsection{Dymnikova  I BH}
For the Dymnikova I metric  we obtain 
\ba
\rho(r)= \frac{ m \ell}{ \pi^2 ( \ell^2+ r^2)^{2} }\, , \quad \quad 
P(r)=  \frac{ m \ell (r^2- 3 \ell^2) }{ 3\pi^2 ( \ell^2+ r^2)^{3} } \, .
\ea
Correspondingly, the e.o.s. is calculated to be, 
\ba
\label{EOS-H}
P(\rho)= \frac{\rho}{3} \Big(  1- 4 \big(\frac{\rho}{\rho_0})^{\frac{1}{2}}\Big) \, ,
\ea
in which $\rho_0\equiv m/ \pi^2 \ell^3 $. 
Near  the center $P\rightarrow -\rho$ while on large distances it reaches 
the relativistic e.o.s with $P\rightarrow \rho/3$ as in Fan-Wang BH. 

Calculating $r_v, r_*$ and $r_m$ we obtain,
\ba
r_v= \ell\, , \quad \quad
r_*=  \sqrt{3} \ell\, , \quad \quad
r_m=  \sqrt{5} \ell  \, .
\ea
As expected,  $r_v < r_* < r_m$.

The sound speed $c_s^2$ is, 
\ba
c_s^2 = \frac{1}{3}- \frac{2}{3}  \big(\frac{\rho}{\rho_0} \big)^{\frac{1}{2}} \, 
= \frac{r^2 - 5 \ell^2}{ 3( r^2 + \ell^2) }  \, ,
\ea
with a root at $r_m$ and the asymptotic value 
$c_s^2 \rightarrow 1/3$ on long distances.

\subsubsection{Dymnikova  II BH}
For the Dymnikova II metric  we obtain, 
\ba
\rho(r)= \frac{ 3m}{4 \pi^2  \ell^3 } e^{-r^3/\ell^3}\, ,
P(r)=  \frac{ 3 m  (r^3- 3 \ell^3) }{ 4 \pi \ell^6  } e^{-r^3/\ell^3} \, .
\ea
Eliminating $r$ from $\rho(r)$ and plugging the result into $P(r)$ the e.o.s. is obtained as follows,  
\ba
\label{EOS-H}
P(\rho)=-\rho \Big[1+  \ln( \frac{\rho}{\rho_0}) \Big]  \, ,
\ea
in which $\rho_0\equiv 3m/ 4 \pi \ell^3 $. 
Near  the center $P\rightarrow -\rho$ while on large distances it has a non-linear e.o.s. with logarithmic running. 

Calculating $r_v, r_*$ and $r_m$ we obtain,
\ba
r_v= (\frac{18}{27})^{1/3} \ell\, , \quad \quad
r_*=   \ell\, , \quad \quad
r_m= 2^{1/3}  \ell  \, .
\ea
As expected,  $r_v < r_* < r_m$.

The sound speed $c_s^2$ is, 
\ba
c_s^2 = -2 -\ln \big(\frac{\rho}{\rho_0} \big) 
= \frac{r^3 - 2 \ell^3}{ \ell^3 }  \, ,
\ea
with a root at $r_m$.  However, curiously, we observe that $c_s^2$ diverges 
on long distances, indicating superluminal propagation.

Looking at the above results we realize that the first four solutions share qualitatively similar behaviours in which the energy density falls off like 
$\rho \sim r^{-3-\alpha}$ with $\alpha >0$. Furthermore, the e.o.s. approaches the linear one $P= w \rho$, with $w= c_s^2\leq 1$.  
On the other hand, the last solution, Dymnikova II metric, shows different behaviours. The main reason is that $\rho(r)$ falls off exponentially in 
Dymnikova II metric so the e.o.s. on long distance behaves like 
$P \sim - \rho \ln( \frac{\rho}{\rho_0})$ as anticipated in section \ref{Reg-BH-cond}.

\begin{figure}[t]
\begin{center}
	\includegraphics[scale=0.5]{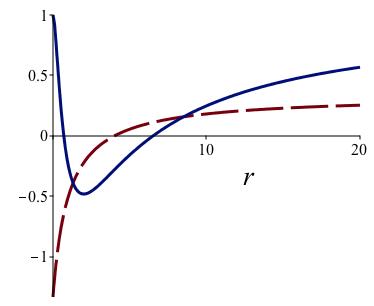}
	\hspace{1.5cm}
	\includegraphics[scale=0.5]{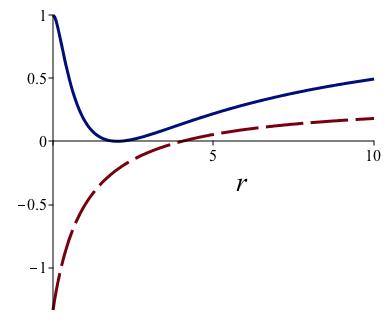}
	\end{center}
\caption{ The joint plot of $A(r)$ (blue solid) and $c_s^2$ (red dashed) for Fan-Wang BH. In the left panel $\ell=1$  and $m=5$ and $ r_- <r_m< r_+$, so the root of $c_s^2$, $r=r_m$, is inside the BH. This is the generic behaviour for large enough value of $m$. 
In the right panel, $\ell=1$ and $m=3.6$ such that $r_m > r_+$. For the given value $\ell=1$, $r_m> r_+$ occurs only in the limited range $ 3.37\leq m\leq 3.85$.
\label{metric-cs-plot}
}
\end{figure}

Before closing this section we comment about the relative positions of the root of $c_s^2$, $r=r_m$, and the roots of metric function $A(r)$, $r=r_\pm$, in the above  five metric solutions. As we have discussed previously,  the sound speed $c_s^2$ becomes negative in some regions  (small $r$ region) which may indicate instabilities. On the other hand, the QNMs perturbations of a few BHs, such as the Bardeen BH, were performed in previous works finding no instability  \cite{Moreno:2002gg, Breton:2005ye, 
Flachi:2012nv, Chaverra:2016ttw, Wu:2018xza, Nomura:2020tpc}. One natural explanation to bypass this apparent paradox is that QNMs describe perturbations for the exterior region so if it is confirmed that the place where $c_s^2$ switches sign is in the interior of BH (i.e. $r_- < r_m< r_+$)
then the two conclusions may be mutually consistent. In 
Fig. \ref{metric-cs-plot} we have presented the joint plot of $A(r)$ (solid curve) and $c_s^2$ (dashed curve) for the Fan-Wang BH. In the left panel 
we observe that $r_- < r_m < r_+$ so the onset of instability occurs inside the BHs. This is generic in the sense that in an extended range of $(\ell, m)$ parameter space the condition $r_- < r_m < r_+$ is satisfied. However, our numerical investigation shows that  it is not universal in the sense that there are small corners of $(\ell, m)$ parameter space that this relation is violated. This is seen in the right panel of Fig. \ref{metric-cs-plot} where the relative value of $(\ell, m)$ are fine-tuned such that $r_m > r_+$.  
Does this signal an instability in this parameter space for QNMs perturbations? This is a technically challenging question which deserve further investigation.

\section{Regular BH Solutions with a Given Equation of State }
\label{generall-eos}

Motivated with the above results, we consider some general forms of 
e.o.s.,
\ba
\label{eos-general}
P= P(\rho)
\ea 
and look for the regular BH solutions.

Our starting point is the energy conservation equation (\ref{energy-eq}) which can be written as follows,
\ba
\label{energy-eq2}
\frac{dr}{r} = -\frac{d \rho}{3 (\rho+ P(\rho))} \, .
\ea
Integrating the above equation, we obtain,
\ba
\label{r-sol}
\frac{r}{\ell} = \exp\Big[ -\frac{1}{3}\int_0^\rho \frac{d \rho'}{ \big(\rho'+ P(\rho') \big) } \Big ] \, ,
\ea
in which $\ell$ is a constant of integration. 
The above equation formally solves $r$ in terms of $\rho$.  Reversing it functionally, we can obtain $\rho=\rho(r)$. Note that this is doable mathematically since the relation between $\rho$ and $r$ is one-to-one due to WEC. After finding $\rho=\rho(r)$, we plug the result in the starting e.o.s. formula in Eq. (\ref{eos-general}) and obtain $P= P(\rho(r))\equiv P(r)$. Finally, with $\rho(r)$ at hand, we obtain the mass function $M(r)$ from the integral (\ref{M(r)}) which yields the metric function $A(r)$.

The above procedure is mathematically straightforward. But the difficulties appear when dealing with the integrations. The first challenge occurs when calculating the integral in the  right hand side of Eq. (\ref{r-sol}). Even after obtaining the integral as $r= r(\rho)$, its reversion may be challenging to find
$\rho=\rho(r)$. Finally, even after obtaining $\rho(r)$, the integral to calculate $M(r)$ from Eq. (\ref{M(r)}) can be challenging too. In summary, we expect that only for limited forms of e.o.s. $P= P(\rho)$ we can solve the system 
analytically to find the configuration solutions $\rho=\rho(r), P= P(r)$ and $A(r)$.

To find suitable equation of state yielding to regular BH solutions, 
the function $f(\rho)\equiv \rho + P(\rho)$ appearing in the denominator of 
Eq. (\ref{energy-eq2}) should meet some certain criteria. First, since $P\rightarrow -\rho$ near the center of BH, then the function $f(\rho)$ should vanish at the point where $\rho$ acquires its maximum value, denoted by $\rho_0$, so $f(\rho_0)=0$. Second, as discussed in section \ref{Reg-BH-cond},  we need $P$ to approach $0^+$  where $\rho\rightarrow 0$ (corresponding to $r\rightarrow \infty$ in configuration space). This imposes the restriction $f(\rho) \rightarrow 0^+$ when $\rho \rightarrow 0$. In particular, it excludes functions of $f(\rho)$ which vanish with higher power $\rho^n$ with $n>1$. Note that $f(\rho)>0$ in the interval $(0, \rho_0)$ to meet the WEC. For a general behaviour of $f(\rho)$ see the right panel of Fig. \ref{P-plot} in which $f(\rho)$ is plotted for the Hayward BH with $\rho_0=1$.

\subsection{Polynomial EOS}
\label{polynomial}

Motivated from the first four examples studied in previous section, we consider the following polynomial form of the e.o.s.,\footnote{See also \cite{Mbonye:2005im} for a somewhat similar e.o.s. using Eq. (\ref{T-lit}) for $T^{\mu}_\nu$. For a linear e.o.s. see also \cite{Hassannejad:2025maw}. }
\ba
\label{P-poly}
P(\rho)= \rho \Big[ w- (1+ w) \big(\frac{\rho}{\rho_0}\big)^n
\Big] \, ,
\ea
in which $w$ plays the role of the linear e.o.s. at $r \rightarrow \infty$ where $\rho \rightarrow 0$. In addition, $\rho_0$ plays the value of $\rho$ at $r=0$. Finally, $n$ is an unspecified free parameter. Demanding that $P \rightarrow 0^+$  as $\rho \rightarrow 0$ we require that $w, n > 0$. With the above form of $P(\rho)$, the function $f(\rho)$ is given by,
\ba
f(\rho)= (1+ w) \rho \Big[ 1- \big(\frac{\rho}{\rho_0}\big)^{n-1} \Big] \, .
\ea
It has the desired property that $f(0)= f(\rho_0)=0$. 

Comparing to cases studied in previous section, the Bardeen BH corresponds to $(w=\frac{2}{3}, n= \frac{2}{5})$, the Hayward BH corresponds to $(w=1, n= \frac{1}{2})$, the Fan-Wang BH corresponds to 
$(w=\frac{1}{3}, n= \frac{1}{4})$ and Dymnikova I BH corresponds to 
$(w=\frac{1}{3}, n= \frac{1}{2})$. 

The sound speed is calculated to be,
\ba
\label{cs-poly}
c_s^2 = w- (1+w) (n+1) \big(\frac{\rho}{\rho_0}\big)^n \, .
\ea
It has a root at 
\ba
\rho_m= \Big[\frac{w}{(1+ w) (n+1)}\Big] ^{1/n} \, ,
\ea
where $P(\rho)$ acquires its maximum,
\ba
P_m = \frac{n w}{1+n}  \Big[\frac{w}{(w+1) n+1}\Big]^{1/n}  \, .
\ea

Plugging the e.o.s. (\ref{P-poly}) into the integral (\ref{r-sol}) yields,
\ba
\ln\big(\frac{r}{\ell} \big) =\frac{1}{3 (1+ w) n} \ln\Big[ \big(\frac{\rho}{\rho_0}\big)^{-n} -1\Big] \, .
\ea
Reversing the above equation, we obtain $\rho(r)$ as follows,
\ba
\label{rhor-poly}
\rho(r)= \rho_0 \Big[  1+ \big( \frac{r}{\ell}\big)^{3 n (1+ w)}
\Big]^{-\frac{1}{n}} \, .
\ea
As expected, the constant $\rho_0$ is the value of $\rho$ at $r=0$. 
In addition, we see that on large $r$, $\rho(r)$ has a power law falls off 
with $\rho(r) \sim r^{-3 (1+w)}$. It falls off faster than $r^{-3}$ for $w >0$ as expected. This in turn guarantees that the total mass $\calM$ given by the integral in Eq. (\ref{M-total}) converges.  {We comment that a similar polynomial profile for $\rho(r)$ as in Eq. (\ref{rhor-poly})
has been studied in \cite{Wang:2026sqr} as well (see also \cite{Wang:2026jvo}). }

With $\rho(r)$ at hand, from the starting e.o.s. (\ref{P-poly}) we can calculate $P(r)$ which we do not present here for brevity. However, we can calculate 
$r_*, r_m$ and $r_v$ from $P(r)$, obtaining 
\ba
r_v = \ell \,  \Big[ \frac{2}{1+ 3 w} \Big]^{\frac{1}{3 (1+ w) n}} \, , \quad \quad
r_*= \ell \, w^{ \frac{-1}{3 (1+ w) n} } \, , \quad \quad
r_m= \ell \,  \Big[\frac{w}{1+ n+ w n}\Big]^{ \frac{-1}{3 (1+ w) n} } \, .
\ea
With some efforts, one can check that the hierarchy $r_v < r_* < r_m$ does hold. 

\begin{figure}[t]
\begin{center}
	\includegraphics[scale=0.45]{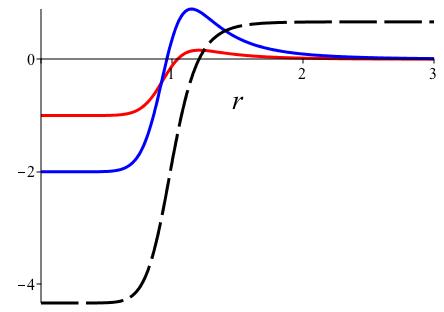}
	\hspace{1.5cm}
	\includegraphics[scale=0.45]{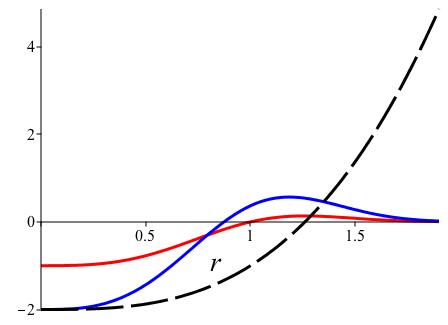}
	\end{center}
\caption{ The plots of $P(r)$ (red), $\rho(r)+ 3 P(r)$ (blue) and $c_s^2$ (dashed black) in the unit $\rho_0=1$.  Left: for the polynomial model Eq. (\ref{P-poly})  with 
$w= \frac{2}{3}$, $n=2$. Right: for the logarithmic model Eq. (\ref{P-rho-log})
with $\beta=1$. In both plots we see $r_v< r_*<r_m$ and $c_s^2 >0 (<0)$ for $r>r_m (r<r_m)$.  
\label{poly-log-plot}
}
\end{figure}

In the left panel of Fig. \ref{poly-log-plot} we have presented $P(r)$, $\rho(r) + 3 P(r)$ and $c_s^2$ for $w=\frac{2}{3}$ and $n=2$. The conclusion 
$r_v < r_* < r_m$ can be seen in the plot.  Also, we see that for $r> r_m (r< r_m)$  $c_s^2>0 (c_s^2 <0)$.

Now, plugging the above solution of $\rho(r)$ into Eq. (\ref{M(r)}) we obtain the mass function $M(r)$ as follows,
\ba
M(r) &=& 4 \pi \rho_0 \ell^3 \int_0^{\ell x} dx\,  x^2  
 \Big[  1+  x^{3 n (1+ w)} \Big]^{-\frac{1}{n}} \nonumber\\
 &=&
\frac{4 \pi \rho_0 \ell^3}{3} F\Big( \frac{1}{n},\, \frac{1}{n (1+ w)},  \, \frac{1+ n (1+ w)}{n (1+ w)}; \, -\big( \frac{r}{\ell}\big)^{3 n (1+ w)} \Big) \, ,
\ea
in which $F(a,b,c; z)$ is the Gauss Hypergeometric function. 

It is promising that an analytical formula for $M(r)$ could be obtained. However, the hypergeometric function is not particularly illuminating to work with. So it would be useful to look for parameter space in which the hypergeometric function is reduced to rational polynomial functions. 
The hypergeometric function given above can be reduced into elementary functions in various cases. Below, we list two categories which yield to interesting results.

\subsubsection{Category 1}
\label{case1}

Inspection shows that when 
\ba
\label{n-w}
n= \frac{w}{k( 1+ w)} \, ,
\ea 
with $k$ being a positive integer, the hypergeometric function reduces to rational polynomial functions.
Starting with $k=1$, we obtain
\ba
\label{Mw-1}
M(r)= \frac{4 \pi \rho_0 r^3 \ell^3}{3}\Big( \ell^{3 w}+r^{3 w} \Big)^{-\frac{1}{w}} \, , \quad 
\quad (n= \frac{w}{1+w}) \, .
\ea
In particular, for $w=2/3, w=1, w=1/3$ we obtain the Bardeen BH, the Hayward BH and the Fan-Wang BH. Curiously, the Dymnikova I BH does not belong to this category. Interestingly, Eq. (\ref{Mw-1}) indicates that an infinite number of regular BH solutions can be obtained which are distinguished by their linear equation of  state parameter $w$.{ Note that the metric (\ref{Mw-1}) was obtained in \cite{Wang:2026sqr} as well}.\footnote{{We are grateful to E. Battista for bring their earlier work \cite{Wang:2026sqr} to our attention after our paper appeared in arXiv.}}. 

The total mass $\calM= M(\infty)$ is obtained to be,
\ba
\label{Minf-1}
\calM= \frac{4 \pi \rho_0 \ell^3}{3}\, , \quad \quad  \quad (n= \frac{w}{1+w}) \, .
\ea
This is a suggestive result!

Next, consider the case $k=2$, yielding
\ba
\label{Mw-2}
M(r)= \frac{4 \pi \rho_0 \ell^3 r^3 }{3(2+w)}
\Big( \ell^{\frac{3w}{2}}+ r^{\frac{3w}{2}} \Big)^{-\frac{2+w}{w}}
\Big[ (2+w)  \ell^{\frac{3w}{2}} + w r^{\frac{3w}{2}} \Big]
 \, , \quad 
\quad \big(n= \frac{w}{2(1+w)} \big) \, .
\ea
The total mass for these cases are given by,
\ba
\label{Minf-2}
\calM= \frac{4 \pi \rho_0 \ell^3}{3(2+ w)} \,   \, , \quad  \quad \quad
\quad \big(n= \frac{w}{2(1+w)} \big) \, .
\ea

Now setting $k=3$, we obtain
\ba
\label{Mw-3}
M(r)= \frac{4 \pi \rho_0 \ell^3 r^3}{3} \frac{ \Big[ 2 w^2r^{2w} + 2 w( 2w+ 3) r^w \ell^w + (9+ 9w + 2 w^2) \ell^2 \Big]  }{(w+3) (2w+3)(\ell^w+ r^w)^{2+ \frac{3}{w}}}  \, , \quad 
\quad \big(n= \frac{w}{3(1+w)} \big) \, .
\ea
The total mass for this category is given by,
\ba
\label{Minf-3}
\calM= \frac{4 \pi  w^3\rho_0  \ell^3}{(w+2) (w+4) (3 w+4)} \,   \, , \quad  \quad 
\quad \big(n= \frac{w}{3(1+w)} \big) \, .
\ea
As can be seen, by increasing the value of $k$, the expression for $M(r)$ becomes more and more complicated so we do not report them for higher values of $k$.

\subsubsection{Category 2}
\label{case2}

The next category where $M(r)$ can be expressed in terms of the elementary function is when $n^{-1}$ is  an integer 
and the following relation between $n$ and $w$ holds,
\ba
\label{n-w}
n= \frac{w}{(k+ \frac{1}{2})( 1+ w)}
\ea 
in which $k$ is an integer (including zero).

Below we list the result for $M(r)$ when $k= \frac{1}{n}-2$ where the results are simple. Note that $n=1$ is excluded so the starting option is $n=\frac{1}{2}$. 
 
With $n=\frac{1}{2}$ and $k=0$, we obtain $w=\frac{1}{3}$ and,  
\ba
M(r) = 2 \pi \rho_0 \ell^3\Big[ \arctan(\frac{r}{\ell}) -\frac{r \ell}{r^2+ \ell^2}
\Big]  \, , \quad \quad (n=\frac{1}{2}, w= \frac{1}{3}) \, ,
\ea
which has the same form as the Dymnikova I solution with the total mass,
\ba
\calM= \pi^2 \rho_0 \ell^3 \, .
\ea

Now suppose $n=\frac{1}{3}$ and $k=1$, we obtain $w=\frac{1}{3}$ and,
\ba
\label{Mw1-cat2}
M(r)= \frac{\pi \rho_0 \ell^3}{2} \Big[ \arctan(\frac{r}{\ell}) + \frac{\ell (r^3- \ell^2 r)}{(r^2+ \ell^2)^2} \Big] \, , \quad \quad (n=\frac{1}{3}, w= 1) \, .
\ea
The total mass for this case is obtained to be,
\ba
\calM= \frac{\pi^2 \rho_0 \ell^3}{4} \, .
\ea

Now suppose $n=\frac{1}{4}$ and $k=2$, we obtain $w=\frac{5}{3}$ and,
\ba
\label{Mw2-cat2}
M(r)= \frac{\pi \rho_0 \ell^3}{4} \Big[ \arctan(\frac{r}{\ell}) + \frac{\ell (r^5 + \frac{8}{3} \ell^2 r^3 - \ell^4 r) \ell)}{(r^2+ \ell^2)^3} \Big] \, , \quad \quad (n=\frac{1}{4},  w=\frac{5}{3})  \, .
\ea
The total mass for this case is obtained to be,
\ba
\calM= \frac{\pi^2 \rho_0 \ell^3}{8} \, .
\ea

As the final example, consider $n=\frac{1}{5}$ and $k=3$, we obtain $w=\frac{7}{3}$, yielding to 
\ba
\label{Mw3-cat2}
M(r)= \frac{5\pi \rho_0 \ell^3}{32} \Big[ \arctan(\frac{r}{\ell}) + \frac{\ell (r^7 + \frac{11}{3} \ell^2 r^5 + \frac{73}{15} \ell^4 r^3  - \ell^6 r) \ell)}{(r^2+ \ell^2)^4} \Big] \, , \quad \quad (n=\frac{1}{5}, w=\frac{7}{3}) \, ,
\ea
with the total mass,
\ba
\calM= \frac{5\pi^2 \rho_0 \ell^3}{64} \, .
\ea

One can continue this line of analysis with higher values of $n^{-1}$ with the condition  $k= \frac{1}{n}-2$. The results would be similar to the above results  but with more complexities as $n^{-1}$ increase.
In addition, one can consider solution with $k= \frac{1}{n}-1$,  
$k= \frac{1}{n}-3$ etc. The results can be expressed in terms of $\arctan$ and fractional powers. However, the results are more complicated than what we reported above so we skip these cases for brevity. Similar results with combinations of power law and $\arctan$ solutions were also obtained in \cite{Wang:2026sqr} from their phenomenological profile of $\rho(r)$.

\subsection{Logarithmic EOS}

As we briefly mentioned in section \ref{Reg-BH-cond}, logarithmic form of e.o.s. appears when $\rho(r)$ falls off exponentially on large distance. Correspondingly, expecting a logarithmic form for e.o.s. is not far fetched. Following by this motivation, let us consider the following form for $P(\rho)$,
\ba
\label{P-rho-log}
P(\rho)= -\rho - \beta \rho \ln\big( \frac{\rho}{\rho_0}  \big) \, ,
\ea
in which $\beta>0$ is a dimensionless constant and $\rho_0$ is another constant which eventually will be the value of $\rho$ at the center of BH.
The above form of $P(\rho)$ is constructed such that $P \rightarrow -\rho$ when $\rho\rightarrow \rho_0$ as required for a dS-type spacetime to regularize the interior of the BH.

The sound speed is obtained to be,
\ba
\label{cs-log}
c_s^2 =-\Big(1+\beta + \beta \ln\big( \frac{\rho}{\rho_0} \big) \Big) \, .
\ea
It has a root at $\rho_m= e^{-\frac{\beta+1}{\beta}}$ where $P$ reaches its maximum,
\ba
P_m= \rho_0 \beta \,  e^{-\frac{\beta+1}{\beta}}  \, .
\ea

Plugging Eq. (\ref{P-rho-log}) into the differential form of the energy conservation Eq. (\ref{energy-eq2}) yields,
\ba
\label{rho-exp}
\rho(r)= \rho_0 \exp\Big[ -\big(\frac{r}{\ell}\big)^{3 \beta} \Big] \, ,
\ea
in which $\ell$ is an integration constant. As expected, $\rho(r)$ has an exponential fall off. Interestingly, the power inside the exponential fall off is controlled by the parameter $\beta$ appearing in 
e.o.s. Eq. (\ref{P-rho-log}). To have a decaying exponential profile, we require $\beta >0$. This exponential form of $\rho(r)$ is known as the Einsato profile, describing  a phenomenological model for the galactic dark matter halo profile \cite{Haud:1986yj, Retana-Montenegro:2012dbd, Konoplya:2025ect, Alencar:2026jwh, Lutfuoglu:2026zel}.

With $\rho(r)$ obtained as in Eq. (\ref{rho-exp}) we can calculate $P(r)$ from the starting equation (\ref{P-rho-log}). Having obtained $P(r)$, we also calculate $r_v, r_*$ and $r_m$, yielding 
\ba
r_v = \Big( \frac{2}{3 \beta} \Big)^{\frac{1}{3 \beta}} \, , \quad \quad
r_*= \beta^{-\frac{1}{3 \beta}}  \, , \quad \quad
r_m=  \Big( \frac{\beta+1}{\beta} \Big)^{\frac{1}{3\beta}} \, .
\ea
With some efforts it can be seen that $r_v< r_*< r_m$.
In the right panel of Fig. \ref{poly-log-plot} we have presented $P(r)$, $\rho(r) + 3 P(r)$ and $c_s^2$ for $\beta=1$. The conclusion 
$r_v < r_* < r_m$ can be seen in the plot.  Also, as  expected, for $r> r_m (r< r_m)$  $c_s^2>0 (c_s^2 <0)$.

Now, plugging $\rho(r)$ obtained in Eq. (\ref{rho-exp}) into the mass function equation (\ref{M(r)}), $M(r)$ is obtained as follows,
\ba
\label{M-exp}
M(r)= \frac{4 \pi \rho_0 \ell^3}{3} \int_0^{(\frac{r}{\ell})^3} dz \exp(- z^\beta) \, .
\ea
The above integrals can be expresses analytically  in terms of the Whittaker functions. Since using Whittaker function may not be particularly insightful, we  try to consider cases where the results can be expressed in terms of elementary functions. Inspection shows that the above integral can be reduced to elementary function when $ \beta^{-1}$ is an integer. In this case, for a given value of $\beta= n^{-1}$ we obtain, 
\ba
M(r) &=&  \frac{4 \pi \rho_0 \ell^3}{3}  n!\Big[ 1-    e^{-z^{3\beta}}   \sum_{k=0}^{n-1} \frac{z^{3\beta  k } }{k!}  
  \Big]  \, , \nonumber\\
  &=&  \frac{4 \pi \rho_0 \ell^3}{3} \Big[ \Gamma(n+1) - n \Gamma(n, z^{3 \beta}) \Big] \,,    \quad \quad  ( z\equiv  \frac{r}{\ell} , ~  n= \frac{1}{\beta})
\ea
in which $\Gamma(n, z)$ is the upper incomplete Gamma function. Correspondingly,  the total mass is given by,
\ba
\calM= \frac{4 \pi \rho_0 \ell^3}{3}  n! \, ,  \quad \quad  (   n= \frac{1}{\beta}).
\ea
It is interesting that $\calM$ as a function of $n$ scales simply as $n!$. Below we consider some examples. 

For $\beta=1$, corresponding to $\rho \propto e^{-z^3}$
we obtain
\ba
M(r) = \frac{4 \pi \rho_0 \ell^3}{3} \big( 1- e^{-z^3} \big)  \, ,
\quad \quad (\beta=1, ~  z = \frac{r}{\ell})  \, .
\ea
This is the solution for Dymnikova II BH.

For $n=2$, corresponding to $\rho \propto e^{-z^{3/2}}$ we obtain
\ba
M(r)= \frac{8 \pi \rho_0 \ell^3}{3}  \Big(1- e^{-z^{3/2}} - z^{3/2} e^{-z^3/2}  \Big) \, , \quad \quad (\beta=\frac{1}{2}, ~  z = \frac{r}{\ell})  \, .
\ea
For $n=3$, corresponding to $\rho \propto e^{-z}$ we obtain,
\ba
M(r)= {4 \pi \rho_0 \ell^3}  \Big(2- (z^2 + 2 z +2) e^{-z}  \Big) \, , \quad \quad (\beta=\frac{1}{3}, ~   z = \frac{r}{\ell})  \, .
\ea

In addition, for $\beta= 2$ (corresponding to a half-integer value $n=1/2$), the result turns out to have a simple form as well,
\ba
M(r)= \frac{2 \pi^{3/2} \rho_0 \ell^3}{3} \mathrm{erf}( z^3) \, , \quad \quad (\beta=2, ~  z = \frac{r}{\ell})  \, ,
\ea
with the total mass given by,
\ba
\calM= \frac{2 \pi \rho_0 \ell^3}{3}  \, , \quad \quad (\beta=2)  .
\ea

While the above results for  various values of $n$ are known in literature under Einasto Halo profile, but we emphasis that we obtained these results from our starting form of $P(\rho)$ given in Eq. (\ref{P-rho-log}). It is surprising that the parameter $\beta$ appearing in $P(\rho)$ plays such non-trivial roles in the solutions. 

\subsection{Trigonometric EOS}
\label{trig-eos}

The last case which we study is  when  $P(\rho)$ is expressed as a trigonometric function with,
\ba
\label{P-trig}
P= -\rho+ \alpha \rho_0 \sin\big( \frac{\beta \rho}{\rho_0} \big) \, ,
\ea 
in which $\alpha$ and $\beta$ are two dimensionless constants while $\rho_0$ is the maximum value of $\rho$ achieved at the center of BH. 
Correspondingly, $f(\rho)= \alpha \rho_0 \sin\big( \frac{\beta \rho}{\rho_0} \big)$. The requirement that 
$f(\rho)$ vanishes linearly like $c \rho$ with $c>1$ for $\rho\rightarrow 0$ is met if $\alpha \beta >1$.  However, the condition that $f(\rho)$ to vanish at  $\rho_0$ imposes the condition,  
\ba
\label{beta-cond}
\beta= n \pi \, ,
\ea
in which $n$ is an integer. Since $\beta $ and $n$ have the same sign, we conclude that $\alpha$ and $n$ have the same sign as well.  Without losing generality,  we assume $\alpha >0$ so $n$ is a positive integer. Furthermore, since  $f(\rho)\ge 0$ in the interval $0\leq \rho \leq \rho_0$ as required by WEC,   then only $n=1$ is allowed. Correspondingly, the condition $\alpha \beta>1$ is translated into $\alpha \pi>1$. 

The sound speed is calculated to be,
\ba
\label{cs-trig}
c_s^2= -1+ \alpha \pi \cos (\frac{\pi \rho}{\rho_0}) \, .
\ea
It has a root at $\rho_m= \frac{1}{\pi}\arccos(\frac{1}{\alpha \pi})$ where $P(\rho)$ assumes its maximum, 
\ba
P_m= - \frac{1}{\pi}\arccos(\frac{1}{\alpha \pi}) + \frac{1}{\pi} \sqrt{\alpha^2\pi^2 -1 } \, .
\ea

Plugging the above form of $P(\rho)$ into the differential equation (\ref{energy-eq2}) we obtain $r(\rho)$. After reversing $r(\rho)$ we obtain the following profile for $\rho(r)$ in configuration space,
\ba
\label{rho-trig}
\rho(r)= \frac{2 \rho_0}{ \pi} \arctan \Big[ \big(\frac{r}{\ell}\big)^{-3 \alpha \pi} \Big] \, ,
\ea
in which $\ell$ is a constant of integration.  
This is an interesting profile. It is reassuring that, since $\alpha \pi >1$, the profile of $\rho(r)$ falls off faster that $r^{-3}$ on large distances. 

\begin{figure}[t]
\begin{center}
	\includegraphics[scale=0.47]{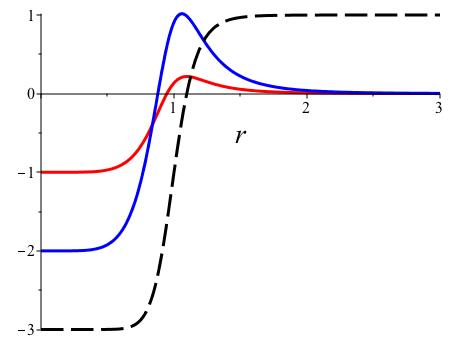}
	\hspace{1.5 cm}
	\includegraphics[scale=0.45]{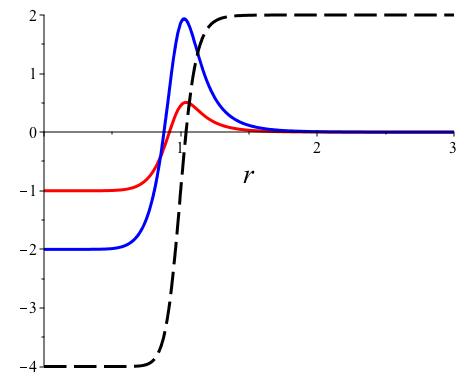}
	\end{center}
\caption{ The plots of $P(r)$ (red), $\rho(r)+ 3 P(r)$ (blue) and $c_s^2$ (dashed black) for the trigonometric e.o.s. (\ref{P-trig})  
for $\alpha \pi=2$ (left) and $\alpha \pi=3$ (right).   In both plots we see $r_v< r_*<r_m$ and $c_s^2 >0 (<0)$ for $r>r_m (r<r_m)$.  In the left panel $c_s<1$ while in the right panel $c_s$ surpasses the unity. 
\label{trig-plot}
}
\end{figure}

With $\rho(r)$ obtained above, one can calculate $P(r)$ from Eq. (\ref{P-trig}) which we do not report here for brevity. From $P(r)$ we obtain,
\ba
r_v= \Big[ \tan\Big(\frac{\pi}{2 (\pi-2)} \Big) \Big]^{-\frac{1}{3 \alpha \pi}} 
\, , \quad \quad
r_*=   \Big[ \tan\Big(\frac{\pi}{2 (\pi-1)} \Big) \Big]^{-\frac{1}{3 \alpha \pi}} \, ,
\quad \quad   
r_m= \Big( \frac{ \alpha \pi+1}{\alpha \pi-1}\Big)^{\frac{1}{6 \alpha \pi}} \, .
\ea
With some analysis one can show that the hierarchy  $r_v < r_* < r_m$ does hold. This conclusion can be seen in Fig. \ref{trig-plot} where $P(r), \rho(r)+ 3 P(r)$ are plotted for $\alpha \pi=2$ and $\alpha \pi=3$.

Plugging the above value of $\rho(r)$ into Eq. (\ref{M(r)}), the mass function is obtained to be,
\ba
\label{M-trig}
M(r)=  8 \rho_0 \ell^3 \int_0^{r/\ell} dx\, x^2 \arctan \big( x^{-3 \alpha \pi} \big) \, .
\ea
The above integral can be expressed in terms of special function Lerch transcendent function. For some special values of $\alpha \pi$ it reduces to elementary functions. Below we report a few examples where the results can be cast in the form of elementary functions. 

For $\alpha \pi =3$, we obtain
\ba
\label{M-alpha=3}
M(r) &=& \frac{2 \rho_0}{3} \ln\Big[ \frac{r^{12}- \ell^6 r^6 +\ell^{12}}{(r^2+\ell^2)^2 (r^4- \ell^2 r^2+\ell^4)^2} \Big]  \nonumber\\
&+& \frac{2 \rho_0}{9} \Big[ \frac{12 r^3}{\ell^3} \arctan\big( \frac{\ell^9}{r^9} \big) 
+ 6 \sqrt3  \arctan\big( \frac{2}{\sqrt3}\frac{r^6}{\ell^6} -\frac{1}{\sqrt3} \big) 
+ \sqrt3 \pi  \Big] \, ,  \quad \quad (\alpha \pi =3) \, .
\ea
It is has a complicated form. But a new feature here is that the argument 
of $\arctan$ contains inverse power of $r$  which is different than the cases obtained in  subsection \ref{case2}. The total mass for this case is obtained to be,
\ba
\calM= \frac{8 \sqrt3 \pi \rho_0 \ell^3}{9} \, ,  \quad \quad (\alpha \pi =3).
\ea

As the second example in this category, consider $\alpha \pi= 2$, yielding,
\ba
\label{M-alpha=2}
M(r) &=& \frac{2 \rho_0}{3} \ln\Big[ \frac{ r^{12}- 2 \sqrt2 \,  r^3 \ell^3 (r^6+ \ell^6)    + 4 \ell^{6} r^6 + \ell^{12} }{ (r^4 +1) (r^8-r^4+1)} \Big]  \\
&+& \frac{2 \rho_0}{3} \Big[ 4 r^3 
\arctan\big( \frac{\ell^6}{r^6} \big) 
+ 2 \sqrt2  \arctan\big( \frac{\sqrt2 r^3}{\ell^3} + 1 \big)  +  2 \sqrt2
\arctan\big( \frac{\sqrt2 r^3}{\ell^3} - 1 \big)
  \Big] \, ,  \quad  (\alpha \pi =2) \, , \nonumber
\ea
with the total mass given by,
\ba
\calM= \frac{4 \sqrt2 \pi \rho_0 \ell^3}{3} \, , \quad  \quad \quad 
 (\alpha \pi =2) .
\ea

\subsection{Subluminal Bound}
\label{super-lum}

As we have shown, the conclusion that $c_s^2$ switches sign at $r=r_m$ is a universal phenomenon in regular BHs, indicating potential instability. However, additional constraint may be imposed on the model parameters 
 if one imposes the subluminal bound $c_s\leq1$. In an effective theory with an unknown UV completion having $c_s>1$ maybe tolerated, but in a true fundamental theory this may indicate pathology. 
With these discussions in mind, let us investigate the constraint $c_s\leq1$ in the models studied above.
Since $c_s^2<0$ for $r< r_m$, we consider the weaker requirement that 
$c_s \leq 1$ for $\rho\rightarrow 0$ corresponding to $r\rightarrow \infty$, so one demands that the IR perturbations to be subluminal.  
 
For the polynomial e.o.s. $c_s^2$ is given in Eq. (\ref{cs-poly}). The condition $c_s\leq 1$  for $\rho \rightarrow 0$ requires only $w\leq1$.
This is the case in all first four models studied in section \ref{reg-BH}. Correspondingly, the condition $w\leq 1$ shall be applied to all models obtained  in subsections \ref{case1} and \ref{case2} as well. 

For the case of logarithmic e.o.s., the sound speed is given in Eq. (\ref{cs-log}). The condition $c_s\leq 1$  is translated into 
\ba
\beta \ln(\frac{\rho}{\rho_0}) \ge -(2 + \beta) \, .
\ea
One can easily see that independent of the value of $\beta$, this condition is violated  when $\rho\rightarrow 0$. This can be seen in the right panel of Fig. \ref{poly-log-plot} where $c_s$ surpasses unity on large distances. 
Correspondingly, the subluminal condition  discards all models, including the Dymnikova II model, in which $\rho(r)$ falls off exponentially as in Eq. (\ref{rho-exp}). This is an unexpected strong limitation. 

Finally, for the case of trigonometric e.o.s., from $c_s^2$ given in 
Eq. (\ref{cs-trig}), the requirement $c_s\leq1$ is cast into, 
\ba
\alpha \pi \cos\big(\pi \frac{\rho}{\rho_0} \big) \leq 2. 
\ea
For $\rho \rightarrow 0$ one obtains the bound $\alpha \pi \leq 2$. Combined with the WEC requirement, we conclude that $1< \alpha \pi \leq 2$. For example, the mass function in Eq. (\ref{M-alpha=3}), corresponding to $\alpha \pi=3$, is excluded while the mass function Eq. (\ref{M-alpha=2}), corresponding to $\alpha \pi=2$, is allowed. This can be seen in 
Fig. \ref{trig-plot}  where $c_s\leq 1$ for $\alpha \pi = 2$ (left panel) while
$c_s$ surpasses unity on long distances for $\alpha \pi =3$ (right panel).

The main goal of this work was to introduce the anisotropic energy momentum tensor such as in Eq. (\ref{T-mu-nu}) and classify the solutions of the metric and the mass function based on their hydrodynamical equation of state $P=P(\rho)$. We believe that this approach is useful as the isotropic part of the energy momentum tensor brings the analysis 
in contact with the standard approaches  in fluid dynamics and thermodynamics. Using this approach one can properly distinguish the important roles that the sound speed $c_s^2$ plays as the criteria for the stability of the setup and the possible occurrence of hydrodynamical instability. More specifically, we have shown that  $c_s^2$ always switches sign in the regular BHs solutions. This is a universal conclusion because, as  we have shown, the construction of the regular BHs solutions require that $P(\rho)$ develops a maximum in the phase space (and correspondingly in the configuration space) so $c_s^2$ switches sign at that point. Whether or not that is physically harmful depends on whether  
 the instability occurs inside the horizon or in the exterior region. 
We have shown that the onset of instability typically occurs inside the horizon so the instability is hidden to the exterior observer. However, this is not universal and depending on model parameters the instability may occur in the exterior region as well. Furthermore, the criteria $c_s \leq 1$ imposes the subluminal requirement. The subluminal condition eliminates some model of regular BHs which may not look like pathological  otherwise.

\section{Summary}

In this work  we have studied the construction of spherically symmetric regular BHs from a given e.o.s. $P= P(\rho)$.
We have assumed that the WEC holds throughout and the energy density falls off fast enough to yield a finite total mass $\calM$. In addition, we have assumed that both $\rho$ and $P$ are finite as necessary conditions for the construction of the regular BH solutions.  During the course of this construction, a number of generic properties emerged which we summarize here. First, as it is well-known, to have a regular BH solution the spacetime near the center should approach a dS background with $P=-\rho$. Furthermore, the SEC should be violated somewhere. Using the energy conservation equation (\ref{energy-eq}) we have demonstrated that the profile of $P(r)$ should be such that it has a root and a maximum and approach $0^+$ on large distances. Correspondingly, the function $f(\rho)= \rho+P$ has two roots at $\rho=0$ and at $\rho=\rho_0$ in which $\rho_0$ is the maximum value of $\rho$ associated to its value at $r=0$ in configuration space. Because of WEC, $f(\rho)>0$ in the interval $(0, \rho_0)$.

We have considered various examples of $P(\rho)$ and obtained the corresponding regular BH solutions. Some of these metric solutions are known in the literature and some are new. The first case which we studied 
was the polynomial e.o.s. given in  Eq. (\ref{P-poly}). Many known regular BHs such as the Bardeen BH, Hayward BH, Fan-Wang BG and Dymnikova I BH belongs to this category. 
The parameter $w$ plays the role of the linear e.o.s. at far distances. The resultant mass function of the metric is given in Eqs. (\ref{Mw-1}),  (\ref{Mw-2})  and  (\ref{Mw-3}) in which the first one is the extension of the  known solutions in the literature. Three other solutions are presented in 
Eqs. (\ref{Mw1-cat2}), (\ref{Mw2-cat2}) and (\ref{Mw3-cat2}) as well while many other new solutions can be constructed via appropriate choosing of the parameters $w$ and $n$. The second case we studied 
was the logarithmic e.o.s. given in Eq. (\ref{P-rho-log}). This type of e.o.s. captures the solutions in which  $\rho(r)$ has exponential fall off such as in Dymnikova II BH or the Einasto profile for the galactic dark matter halo. The third case was the trigonometric e.o.s. given in Eq. (\ref{P-trig}). This leads to the new solutions of the mass function  such as in Eqs. (\ref{M-alpha=3}) and  (\ref{M-alpha=2}) and many other new solutions which we did not report  for brevity. 

It is shown that the hierarchy $r_v < r_* <r_m$ does hold. Furthermore since 
$\partial P/\partial \rho=0$ at $\rho_m=\rho(r_m)$, then $c_s^2$ changes sign at $r_m$.  This is a universal phenomena afflicting all models of regular BHs. The key to the origin of this effect is the property of $P(r)$ that it should have a maximum and approach $0^+$ at large distances. Whether or not  a negative $c_s^2$ indicates an instability requires further investigation. Our numerical analysis show that  in 
the extended region of parameter space  $r_- < r_m < r_+$  so 
 $c_s^2 >0$ for the exterior of BH and the corresponding QNMs may be stable. This conclusion requires careful investigation, specially in the limited
 corner of parameter space where  $r_m > r_+$.  In addition, imposing the subluminal bound $c_s\leq 1$ puts strong constraints on the model parameters. In particular, models in which $\rho(r)$ has an exponential 
 fall off as in Eq. (\ref{rho-exp}) are excluded. 

We have advocated that the anisotropic energy density in the form Eq. (\ref{T-general}) or (\ref{T-mu-nu}) is better suited for the specification of e.o.s. as it naturally separates the isotropic pressure and the anisotropic stress. For example, as studied in  Appendix \ref{Maxwell-example}, 
one can show that for a relativistic Maxwell field it yields the correct linear e.o.s. $P=\rho/{3}$ while the common formulation represented by Eq. (\ref{T-lit}) yields the e.o.s. $P_r=-\rho, P_t = + \rho$ which may not be interpreted naturally from a relativistic field theory.  

One shortcoming of our analysis is that the proposed e.o.s. $P=P(\rho)$ was mostly considered from a phenomenological and hydrodynamic point of view. We did not provide a dynamical mechanism, i.e. field theory realization of the proposed e.o.s. For example, it is known that many regular BH solutions, including the known models studied in section \ref{reg-hyd}, can be obtained from non-linear electrodynamics with the Lagrangian of the form $L=L(F)$. However, a shortcoming associated with the non-linear electrodynamics realization of regular BHs is that these Lagrangians do not converge to linear Maxwell theory in the low energy limit. Having said this, it is an important question to examine what field theory (if any) can yield the desired e.o.s. studied in our work for regular BHs construction.

\vspace{0.7cm}

{\bf Acknowledgments:}  We thank Hideki Maeda for insightful comments and correspondences. We acknowledge Maple Software for computational analysis  and  DeepSeek for assistance. This work is supported by the INSF of Iran under the grant number 4045105.
  


\appendix 
\section{An Example of Anisotropic Fluid}
\label{Maxwell-example}

Here we present the free Maxwell theory as an example of the anisotropic fluid where the construction of $T_{\mu \nu}$ in the form (\ref{T-mu-nu}) 
is presented. As is well-known the free Maxwell  theory yields the Reissner-Nordstrom (RN) BH which is singular at the center. 

Starting with the action, 
\ba
S= -\frac{1}{4} \int d^{4} x \sqrt{-g}  F_{\mu \nu} F^{\mu \nu} \, ,
\ea
the corresponding energy momentum tensor as usual  is given by, 
\ba
\label{T-em}
T^{\mu}_{\nu} =    F^{\mu}_{~ \kappa} F_{\nu}^{~ \kappa}   -\frac{1}{4} \delta^{\mu}_{\nu} {\cal F}   \, ,
\ea
in which we have defined  ${\cal F} \equiv F_{\mu \nu} F^{\mu \nu} $. 

Consider an electrically charged BH with $F_{01} \neq 0$. From the Maxwell equation $\nabla_\mu F^{\mu \nu}=0$, we obtain $F_{01} = \bar Q/r^2$ in which $\bar Q$ is the constant electric charge of the system. 

Plugging the above solution of $F_{\mu \nu}$,  the energy momentum tensor is obtained to be 
\ba
 {T^{\mu}_{\nu}} = \frac{{\cal F}}{ 4} \mathrm{diag} 
 \left(  1, 1, -1, -1  \right) \, ,
\ea
 in which,
 \ba
 {\cal F} = \frac{-2  \bar Q^2}{r^4} \, .
 \ea

Comparing the above energy momentum tensor with our form of $T^{\mu}_{ \nu}$ given in Eq. (\ref{T-mu-nu}), we can  construct $\rho$, $P, \Pi$ as follows, 
\ba
\label{rho-Max}
\rho = -\frac{{\cal F}}{ 4 } = \frac{\bar Q^2}{2 r^4 }
, \quad  P = -\frac{{\cal F}}{ 12  }, \quad
\Pi =  -\frac{{\cal F}}{ 6  } \, .
\ea
One can easily check that  $2 \Pi = \rho + P$ as a fundamental requirement for the consistency of the solution with $A(r)=B(r)$ 
in the general metric Eq. (\ref{metric-BH}). Also, the e.o.s. is obtained to be,
\ba
w= \frac{P}{\rho}= \frac{1}{3} \, .
\ea
This is consistent with the usual expectation that for the photon particles 
$w=\frac{1}{3}$. The fact that we have anisotropies should not alter this 
fundamental property of the free Maxwell theory. On the other hand, if one uses the usual form of $T^\mu_\nu$ as in Eq. (\ref{T-lit}) one obtains the e.o.s. $P_r= -\rho$ which naively represents a dS type of matter. But this is misleading as for the remaining component we obtain the e.o.s. $P_t= +\rho$. The way out of this paradoxical situation is to decompose $T^\mu_\nu$ into its isotropic and anisotropic components and define the proper e.o.s. for the isotropic sector, i.e. $P= P(\rho)$ as we did in this work.

Finally, the RN metric function is obtained to be, 
\ba
A(r) = B(r) = 1- \frac{2 M}{r} +  \frac{ Q^2}{ r^2 }  \,  , 
\quad (Q^2\equiv 4 \pi  G \bar Q^2)  \, .
\ea

%


\bibliography{BH-EOS}{}

\bibliographystyle{JHEP}


\end{document}